\documentclass[aps,twocolumn,showpacs,preprintnumbers,prb]{revtex4}
\usepackage{graphicx}
\usepackage{color}
\begin{document}

\title{{\it Ab initio} calculation of low-energy collective charge-density excitations in
$\textrm{MgB}_2$ }

\author{V. M. Silkin,$^{1,2,3}$ A. Balassis,$^{2,4}$  P. M.
Echenique,$^{1,2}$ and E. V. Chulkov$^{1,2}$}

\affiliation{ $^1$Depto. de F\'{\i}sica de Materiales and Centro
Mixto CSIC--UPV/EHU, Facultad de Ciencias Qu\'{\i}micas, Universidad
del Pa\'{\i}s Vasco, Apdo. 1072, 20080 San Sebasti\'an, Basque
Country, Spain\\
$^2$Donostia International Physics Center (DIPC), P. Manuel
Lardizabal 4, 20018 San Sebasti\'an, Basque Country, Spain\\
$^3$IKERBASQUE, Basque Foundation for Science, 48011, Bilbao, Spain \\
$^4$Physics Department, Fordham
University, 441 East Fordham Road, Bronx NY 10458-5198, USA }

\begin{abstract}
We present {\it ab-initio} time-dependent density-functional theory
calculation results for low-energy  collective electron excitations
in $\textrm{MgB}_2$. The existence of a long-lived  collective
excitation corresponding to coherent charge density fluctuations
between the boron $\sigma$- and $\pi$- bands ($\sigma\pi$ mode) is
demonstrated. This mode has a sine-like oscillating dispersion for
energies below 0.5 eV. At even lower energy we find another
collective mode ($\sigma\sigma$ mode). We show the strong impact of
local-field effects on dielectric functions in MgB$_2$. These
effects account for the long q-range behavior of the modes. We
discuss the physics that these collective excitations bring to the
energy region typical for lattice vibrations.
\end{abstract}

\date{\today}


\maketitle

After several years of intense experimental and theoretical study of
the origin of unusually high temperature
superconductivity\cite{nanan01} in MgB$_2$, nowadays there is a
general belief that the superconductivity in MgB$_2$ is
phonon-mediated with multiple gaps and an exceptional role played by
$E_{2g}$ phonon mode in strong coupling to electrons in the
tubular-shaped $\sigma$ bands at the Fermi level, $E_F$. However,
despite the tremendous progress achieved in understanding of
superconductivity in MgB$_2$, many of its fundamental properties
related to phonons and electron-phonon ($e-ph$) coupling are still
puzzling and remain under intense debate. Thus such important issues
as, e.g., the reduced isotope
effect,\cite{bulaprl01,hicln01,sipacm08} the nature of the strong
broad continuum in Raman spectra, the unusually broad linewidth of
the $E_{2g}$ phonon mode seen in Raman
spectra,\cite{boheprl01,gostprb01,quleprl02} the inconsistency
between the results of Raman scattering and x-ray
measurements,\cite{shcaprl03,baucprl04,baucprb07} and the role of
non-adiabatic
effects\cite{shcaprl03,caprb06,calapc07,dacaprb07,salaprl08} have
not yet found satisfactory explanations.

As for the electronic part of the $e-ph$ picture of
superconductivity in MgB$_2$, it has received substantially less
attention than the phonon part. In particular, in all evaluations of
the $e-ph$ interaction the adiabatic approximation has been used,
and even in {\it ab initio} calculations\cite{flprprl05} of $T_c$
the dynamical Coulomb interaction has been considered in its static
form. Here we present a detailed {\it ab initio} study of the
low-energy dynamical electronic properties of MgB$_2$. We
demonstrate that the strongly anisotropic electronic structure of
MgB$_2$ characterized by boron quasi two-dimensional $\sigma$ and
three-dimensional $\pi$ bands\cite{komaprl01,anpiprl01} leads to
remarkable low-energy dielectric response in this compound:
collective modes with a peculiar sine-like oscillating dispersion
appear  in the 0-0.5 eV energy range. This brings interesting
physics to the energy region which was thought to be entirely
dominated by lattice vibrations.

Our approach is based on time-dependent density functional
theory\cite{pegoprl96} where the non-local dynamical
density-response function, $\chi$, determines an induced charge
density, $\rho$, in the electronic system caused by an external
potential, $\upsilon_{\rm ext}$, according to
\begin{equation}\label{Induced_density}
\rho({\bf r},t) = \int \chi({\bf r},t;{\bf r}',t')\,\upsilon_{\rm
ext}({\bf r}',t') \,{\rm d}{\bf r}' {\rm d}t'.
\end{equation}
$\chi$ is obtained from the integral equation $\chi = \chi^o +
\chi^o(\upsilon+K_{\rm xc})\chi$, where $\chi^o({\bf r},t;{\bf
r}',t')$ is the response function of non-interacting electron
system, $\upsilon({\bf r}-{\bf r}')$ is the bare Coulomb potential,
and $K_{\rm xc}$ accounts for dynamical exchange-correlation
effects. In terms of Fourier-transformed quantities the imaginary
part of the matrix $\chi^o_{{\bf G}{\bf G}'}({\bf q},\omega)$ can be
evaluated using equation $S^0_{{\bf G}{\bf G}'}({\bf
q},\omega)=-\frac{1}{\pi}{\rm sgn}(\omega){\rm Im}[\chi^o_{{\bf
G}{\bf G}'}({\bf q},\omega)]$,\cite{arguprb94} with the spectral
function $S^0_{{\bf G}{\bf G}'}({\bf q},\omega)$ defined as
\begin{eqnarray}\label{spectral_function}
S^0_{{\bf G}{\bf G}'}({\bf q},\omega)= \frac{2}{\Omega} \sum^{\rm
BZ}_{\bf k} \sum_{n}^{\rm occ} \sum_{n'}^{\rm unocc}
\delta(\varepsilon_{n{\bf k}}-\varepsilon_{n'{\bf k}+{\bf
q}}+\omega) \nonumber\\
\times \langle\psi_{n{\bf k}}|e^{-{\rm
i}({\bf q}+{\bf G})\cdot{\bf
r}}|\psi_{n'{\bf k}+{\bf q}}\rangle \nonumber\\
\times\langle\psi_{n'{\bf k}+{\bf q}}|e^{{\rm i}({\bf q}+{\bf
G}')\cdot{\bf r}}|\psi_{n{\bf k}}\rangle
\end{eqnarray}
and Re$[\chi^o_{{\bf G}{\bf G}'}({\bf q},\omega)]$ is evaluated via
the Kramers-Kronig relation using an energy cutoff of 50 eV. Here
the ${\bf G}$'s are the reciprocal lattice vectors, $n$,$n'$ are
band indices, the wave vectors ${\bf k}$ and ${\bf q}$ are in the
first Brillouin zone (BZ), the factor 2 accounts for the spin, and
$\Omega$ is the normalization volume. In practice, we replace the
$\delta$-function in Eq. (\ref{spectral_function}) by a Gaussian
$\frac{1}{\sqrt{\pi}\gamma}e^{-\omega^2/\gamma^2}$ with a very small
broadening parameter $\gamma=10$ meV. The sum over ${\bf k}$ was
performed on a $108\times108\times360$ grid. The use of such a fine
${\bf k}$-mesh sampling is crucial to achieve convergence of
dielectric properties in MgB$_2$ at low energies. The one-particle
energies $\epsilon_{n{\bf k}}$ and wave functions $\psi_{n{\bf k}}$
are obtained as self-consistent solutions of Kohn-Sham equations
using norm-conserving pseudopotentials\cite{trmaprb91} and the
exchange-correlation potential of Ref. \onlinecite{cealprl80}. In
order to elucidate the role of exchange-correlation effects in the
density-response function, we performed calculations of $\chi$ using
two forms of the many-body kernel $K_{\rm xc}$, namely the
random-phase approximation (RPA) ($K_{\rm xc}$=0), and an adiabatic
extension of the local-density approximation (TDLDA).\cite{grdo96}

\begin{figure}
\includegraphics[scale=1.6,angle=270]{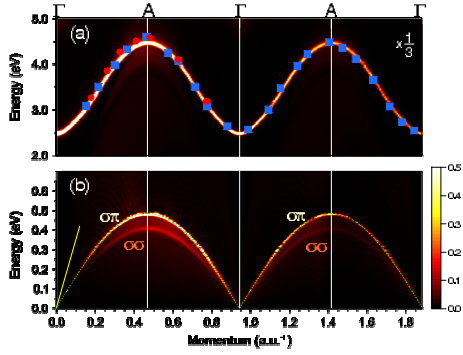}
\caption{(color). Calculated  $-{\rm
Im}[{\epsilon^{-1}}(q_c,\omega)]$ versus $\omega$ and $q_{c}$ in two
energy regions. Calculations include local-field effects and the RPA
kernel. In (a) the circles and squares mark the energy-loss peak
positions measured in x-ray scattering experiments of Refs.
\onlinecite{gasoprb05} and \onlinecite{cachprl06}, respectively. In
(b) the dispersion of the upper sharp $\sigma\pi$ plasmon peak is
described by
$\omega_{\sigma\pi}=V_{\sigma\pi}\cdot|\sin(\frac{c}{2}q_c)|$ (green
dashed line) with $V_{\sigma\pi}=0.48$ eV. The lower broad feature
with similar sine-like dispersion corresponds to the $\sigma\sigma$
plasmon. Yellow solid line presents acoustic plasmon dispersion
according to Ref. \onlinecite{voancm01}. } \label{ALP_dispersion}
\end{figure}

The calculated loss function, $-{\rm Im}[\epsilon^{-1}({\bf
q},\omega)]$,\cite{note_1} directly probed in inelastic scattering
experiments is presented in Fig. \ref{ALP_dispersion} as a function
of the momentum transfer ${\bf q}$ along the $c^*$ axis. In the
upper energy range (Fig. \ref{ALP_dispersion}a) it is dominated by a
well-defined collective mode dispersing in the 2.5-4.5 eV energy
range in excellent agreement with x-ray
experiments.\cite{gasoprb05,cachprl06} The existence of this mode
for momenta in the first BZ was demonstrated in Refs.
\onlinecite{zhsiprb01} and \onlinecite{kupiprl02}, whereas its
cosine-like oscillating dispersion in higher BZ's was recently
discovered in a joint experimental-theoretical
study.\cite{cachprl06} This mode produces a strong impact on
electrodynamical and optical
properties\cite{kaleprl06,gukuprb06,kupc07} of MgB$_2$, however, it
has no relevance to superconductivity as it affects neither the
dynamically screened Coulomb interaction at low energies nor phonon
dispersion.\cite{kupiprl02}

\begin{figure}
\includegraphics[scale=.42,angle=0]{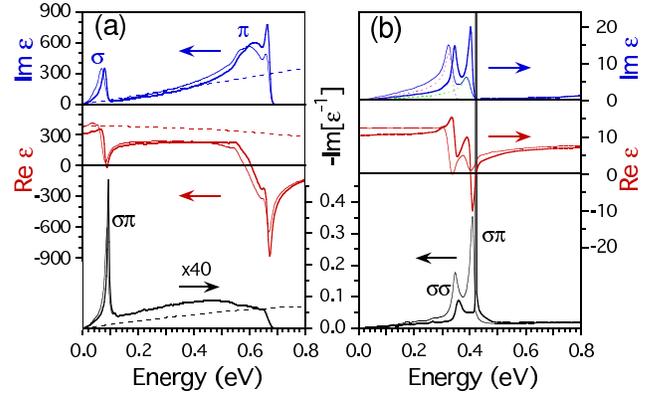}
\caption{(color).  Real (red lines) and imaginary (blue lines) parts
of the dielectric function, $\epsilon(q_c,\omega)$, and the
corresponding loss function, $-{\rm Im}[\epsilon^{-1}(q_c,\omega)]$,
(black lines) for (a) $q_c=0.055$ a.u.$^{-1}$ and (b) $q_c=0.322$
a.u.$^{-1}$ evaluated with the use of the RPA kernel. Thick (thin)
lines stand for results with (without) inclusion of local-field
effects. In (a) the dashed lines are the free-electron gas results
for $r_s=1.8$ a.u. In (b) thin dotted lines show contributions to
${\rm Im}\epsilon$ due to intraband transitions within the
$\sigma_1$ (green) and $\sigma_2$ (purple) bands. }
\label{Eps_q_RPA}
\end{figure}

\begin{figure}
\includegraphics[scale=1.75,angle=270]{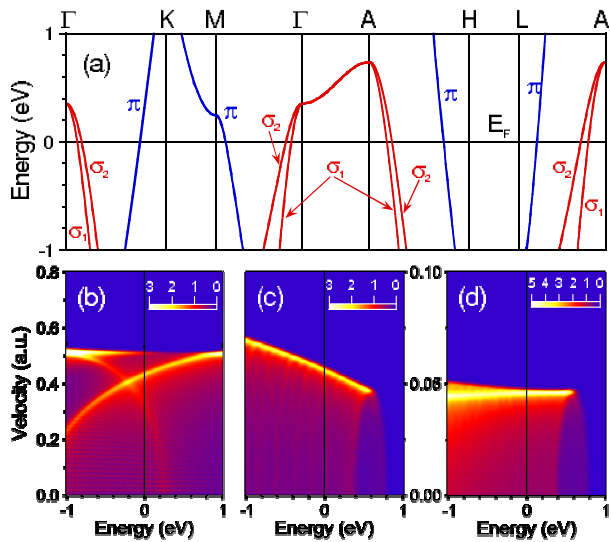}
\caption{(color). (a) Energy band structure of MgB$_2$ along some
symmetry directions of the Brillouin zone in the vicinity of the
Fermi level, $E_F$. Maps of the density of states in the $\pi$ (b),
$\sigma_1$ (c), and $\sigma_2$ (d) bands versus their energy and
group velocity component along the $c^*$ axis. Note the different
velocity range of (b) in comparison with (c) and (d). }
\label{Band_structure+velocity}
\end{figure}

Figure \ref{ALP_dispersion}b shows the calculated loss function in
the low-energy domain that is our main finding. At all $q_c$ it is
dominated by a sharp $``\sigma\pi"$ peak with a sine-like
dispersion. Figure \ref{Eps_q_RPA} shows the dielectric and loss
functions at $q_c=0.055$ a.u.$^{-1}$ and $q_c=0.322$ a.u.$^{-1}$.
One can see how the presence of two types of carriers (in the
$\sigma$ and $\pi$ bands) around $E_F$ (see Fig.
\ref{Band_structure+velocity}a) with a large difference in the
perpendicular component of the group velocity, $\upsilon^{\perp}$,
(compare maximal $\upsilon^{\perp}$ in the $\pi$ bands (Fig.
\ref{Band_structure+velocity}b) with that in the $\sigma$ bands
(Figs. \ref{Band_structure+velocity}c and
\ref{Band_structure+velocity}d)) produces in Im$\epsilon$ a
structure consisting of two main peaks (Fig. \ref{Eps_q_RPA}a).
Thus, while the faster $\pi$ carriers give rise to a broad structure
from 0 to 0.68 eV in Fig. \ref{Eps_q_RPA}a with the main peak at the
upper-energy side, the $\sigma$ carriers which are moving more
slowly in the $c^*$ direction produce a sharp peak in the low-energy
part of Im$\epsilon$. The reason of this sharpness resides in the
fact that within both $\sigma$ bands, the number of states with
maximal $\upsilon^{\perp}$ is greatly enhanced as seen in Figs.
\ref{Band_structure+velocity}c-\ref{Band_structure+velocity}d.
Therefore, the number of intra-band transitions involving these fast
states is large which strongly enhances the $\sigma$ peak in
Im$\epsilon$ at the higher energies. In turn, this causes a dramatic
drop in ${\rm Re}\epsilon$ at nearly the same energy which, combined
with the presence of a local minimum around $\omega=0.1$ eV in ${\rm
Im}\epsilon$, produces a well-defined sharp peak in $-{\rm
Im}[\epsilon^{-1}]$ corresponding to charge density fluctuations
between the $\sigma$ and $\pi$ bands, a $\sigma\pi$ mode. Despite
the presence of a non-vanishing value of Im$\epsilon$ at the
$\omega$ where the $\sigma\pi$ peak appears, its intrinsic width is
extremely small, being in the meV range (which corresponds to a
lifetime of few hundreds of femtoseconds), and is almost entirely
determined by an extrinsic broadening parameter $\gamma$.

Since at small $q_c$ the two main peaks in Im$\epsilon$ disperse
linearly with momentum according to their maximal perpendicular
Fermi velocity components, Im$\epsilon \propto\upsilon_{F,{\rm
max}}^{\perp\sigma(\pi)}\cdot q_c$, the dispersion of the
$\sigma\pi$ mode (linked to the upper side of the continuum for
electron-hole excitations within the $\sigma$ bands) is also linear
in $q_c$ and its group velocity $\upsilon_{\sigma\pi}$ is close to
$\upsilon_{F,\rm max}^{\perp\sigma}$.\cite{note_3} This corresponds
to the acoustic plasmon proposed to exist in MgB$_2$ on the basis of
a model tight-binding calculation.\cite{voancm01} The concept of an
acoustic plasmon goes back to Pines\cite{picjp56} who showed that it
can occur in a two-component electron plasma consisting of slow and
fast carriers. The fast carriers (within the $\pi$ bands in MgB$_2$)
can act to screen the repulsion between slow carriers (within the
$\sigma$ bands in MgB$_2$) resulting in the appearance of a plasmon
mode with a peculiar sound-like dispersion. After the recognition of
the role in the acoustic plasmon as a possible mechanism for
superconductivity in transition metals,\cite{frjpc68} it has been
regularly evoked for explaining superconductivity in materials with
unusually high $T_c$. Nevertheless, up to now the issue remains
controversial since the very existence of such a collective mode in
metals has been confirmed neither experimentally nor by {\it ab
initio} calculations. Only recently an acoustic-like plasmon was
observed in the electron energy-loss measurements at a metal surface
in excellent agreement with the {\it ab initio}
prediction,\cite{dipon07} thus greatly increasing our confidence in
the present results.

With increasing $q_c$ the $\sigma$ peak in Im$\epsilon$, and
consequently the $\sigma\pi$ peak in $-{\rm Im}[\epsilon^{-1}]$,
starts to split into two peaks (Figs. \ref{ALP_dispersion}b and
\ref{Eps_q_RPA}b) since for holes, $\upsilon^{\perp\sigma_1}_{\rm
max}$ exceeds $\upsilon^{\perp\sigma_2}_{\rm max}$ by more than
20$\%$ (Figs. \ref{Band_structure+velocity}(c) and
\ref{Band_structure+velocity}(d)). In this case the
higher(lower)-energy plasmon peak in $-{\rm Im}[\epsilon^{-1}]$
corresponds to the charge fluctuations between $\sigma_1$ and $\pi$
($\sigma_1$ and $\sigma_2$) bands. Whereas the $\sigma\pi$ plasmon
continues to be long-lived, the lower-energy $\sigma\sigma$ plasmon
has a significantly shorter lifetime due to more efficient
scattering within the $\sigma_1$ band. We expect the two separate
modes to exist at any $q_c$, however, when the calculated broadening
$\gamma$ exceeds the energy difference between these modes only a
single $\sigma\pi$ peak arises in $-{\rm Im}[\epsilon^{-1}]$ (Fig.
\ref{Eps_q_RPA}a).\cite{note_2}

In Fig. \ref{ALP_dispersion}b one can see how with increasing $q_c$
both of these plasmon modes reach maximum energy at the A point and,
as momentum increases further, their dispersions change from
positive to negative and approach $\omega=0$ at the $\Gamma$ point
in the second BZ, in striking contrast to results of Ref.
\onlinecite{voancm01}. The origin of this periodic behavior resides
in the fact that strong local-field effects in MgB$_2$ feed the
strength from the small $q_c$ modes into the charge density
fluctuations at large $(q_c+G_z)$ in a similar fashion as occurs in
the case of the higher energy mode.\cite{cachprl06} These sine-like
dispersing modes continue to have strength at momenta in subsequent
BZ's and this is a direct consequence of a layered MgB$_2$
structure. Additionally, as seen in Fig. \ref{Eps_q_RPA}, the
local-field effects lead to a blue shift of the $\sigma\sigma$ and
$\sigma\pi$ energy and make the $\sigma\pi$ peak more narrow as it
occurs for $\omega$ corresponding to smaller Im$\epsilon$. We also
analyzed the role of exchange-correlation effects beyond the RPA and
found that the RPA picture is essentially sufficient for the
description of low-energy collective excitations in MgB$_2$. The
inclusion of the TDLDA kernel in the calculation of
$\chi(q_c,\omega)$ leads to only a small (few percent) downward
shift of the plasmon mode dispersions with a slight reduction of
their lifetimes, i.e., in part compensating the local-field effects.

\begin{figure}
\includegraphics[scale=.5,angle=0]{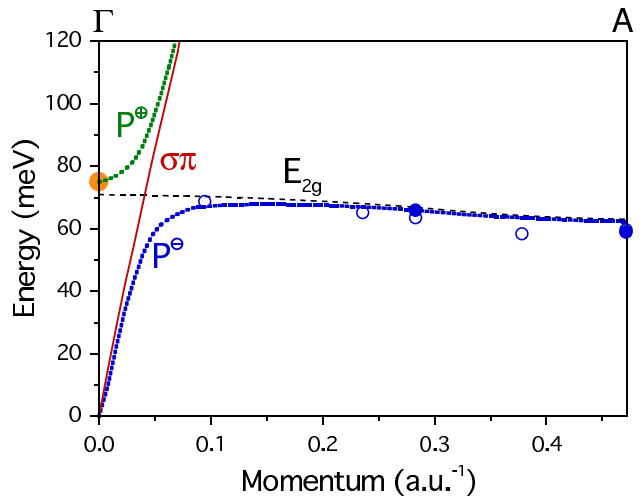}
\caption{(color online). The low-energy part of the bare $\sigma\pi$
plasmon (red solid line) along with the calculated\cite{yiguprl01}
bare $E_{2g}$-phonon (dashed line) and measured peak positions
(open\cite{shcaprl03} and filled\cite{dacaprb07} circles). The
position of the Raman spectrum peak is shown by an orange filled
circle at $q_c=0$. Dispersions of the hybrid ``$\sigma\pi-E_{2g}$"
modes, $P^{\oplus}$ and $P^{\ominus}$, are shown by green and blue
thick dotted lines, respectively. In the evaluation of these curves
we use a ``$\sigma\pi-E_{2g}$"  coupling parameter $\Delta$=17.3 meV
in order to place the position of the $P^{\oplus}$ mode at $q_c=0$
in agreement with the Raman
measurements.\cite{boheprl01,gostprb01,quleprl02} }
\label{ALP+phonons}
\end{figure}

The $\sigma\pi$ plasmon can have deep impact on both
low-energy-electron and phonon dynamics in MgB$_2$. Our calculation
reveals dramatic modification of the dynamical Coulomb interaction
in the energy range vital for superconductivity. In Fig.
\ref{Eps_q_RPA} one can see that in the neighborhood of the
$\sigma\pi$ plasmon energy the calculated Re$[\epsilon(q_c,\omega)]$
differs dramatically from static $\epsilon(q_c,\omega=0)$ as well as
from the free electron gas result. In particular, in a certain
momentum-energy phase space region, Re$\epsilon$ is even negative
leading to overall reduction of Coulomb repulsion between carriers.
This could explain in part the observed reduced isotope effect in
MgB$_2$.\cite{hicln01,bulaprl01,sipacm08} It would be of great
interest  to quantify this effect within the {\it ab initio}
approach.\cite{flprprl05} Note, that the $\sigma\pi$ plasmon does
not influence inter-band $\sigma$-$\pi$ and intra-band $\sigma$
scattering due to the phase-space restrictions, although could have
some effect in the intra-band $\pi$ one.

The $\sigma\pi$ plasmon can dramatically affect the dispersion of
the optical phonon modes in the ``pathological" region of small
${\bf q}$,\cite{dacaprb07} offering an unexpected explanation of
widely-discussed discrepancies between the Raman and x-ray
measurements of the phonon structure in MgB$_2$. The $\sim77$ meV
peak observed in Raman-scattering experiments and commonly
attributed to the boron $E_{2g}$ phonon mode is always strongly
renormalized in comparison with a bare phonon dispersion, whereas
the x-ray experiments performed for large $q_c$ do not see any
appreciable renormalization. In Fig. \ref{ALP+phonons} we show the
calculated dispersion of the bare $\sigma\pi$ mode along with the
bare $E_{2g}$ phonon mode\cite{yiguprl01} compared to the
experimental data.\cite{shcaprl03,dacaprb07} One can see how the two
bare curves cross each other close to the $\Gamma$ point. Due to its
localization in the boron plane the $\sigma\pi$ plasmon strongly
interacts with the boron $E_{2g}$ phonon mode resulting in strong
hybridization of these boson modes. The result shown by green and
blue dashed lines in Fig. \ref{ALP+phonons} demonstrates that the
mode seen in the Raman experiments is indeed a strongly hybridized
``$\sigma\pi$-$E_{2g}$" $P^{\oplus}$ mode. Note, that the finite
electron lifetime effect\cite{caprb06,salaprl08} can also contribute
to the upward shift of this mode. Additionally, a steep $\sigma\pi$
plasmon dispersion at small momenta might explain the existence of
the strong unstructured background (commonly mentioned as being of
unknown electronic origin) observed in the Raman experiments. This
is corroborated by the fact that, e.g., in AlB$_2$, where we do not
expect the existence of such a plasmon, this background does not
appear.\cite{boheprl01,reboprl02}

In conclusion, our detailed {\it ab initio} calculation demonstrates
the existence in MgB$_2$ in the 0-0.5 eV energy range of hitherto
unknown long-lived collective mode that corresponds to coherent
charge fluctuations between the boron $\sigma-$ and $\pi-$ bands
($\sigma\pi$ mode) with striking periodic sine-like dispersion. This
mode shows an acoustic-like behavior at small momenta where it can
strongly interact with optical phonons. Additionally we find at
slightly lower energy a more strongly damped mode that corresponds
to charge fluctuations between two different $\sigma$ bands
($\sigma\sigma$ mode). Both these modes have profound impact on the
low-energy dynamical Coulomb interaction which should be explicitly
taken into account in {\it ab initio} theories of superconductivity
to have the predictive power.

We acknowledge partial support from the University of the Basque
Country (Grant No. GIC07IT36607), the Departamento de Educaci\'on
del Gobierno Vasco, and the Spanish Ministerio de Ciencia y
Tecnolog\'{\i}a (MCyT) (Grant No. FIS200766711C0101).


\begin{thebibliography}{99}
\bibitem{note_3} In the present numerical calculations this has been
demonstrated down to the energy $\omega\sim25$ meV.
\end{thebibliography}
\end{document}